# Mössbauer study of Eu$_{0.57}$Ca$_{0.43}$Fe$_2$As$_2$ and Eu$_{0.73}$Ca$_{0.27}$(Fe$_{0.87}$Co$_{0.13}$)$_2$As$_2$: A comparison to '122' iron-based superconductors parent compounds EuFe$_2$As$_2$ and CaFe$_2$As$_2$

K. Komędera[1], A. Błachowski[1], K. Ruebenbauer[1*], J. Żukrowski[2], S. M. Dubiel[3], L. M. Tran[4], M. Babij[4], and Z. Bukowski[4]

[1]Mössbauer Spectroscopy Laboratory, Institute of Physics, Pedagogical University
*PL-30-084 Kraków, ul. Podchorążych 2, Poland*

[2]Academic Centre for Materials and Nanotechnology, AGH University of Science and Technology
*PL-30-059 Kraków, Av. A. Mickiewicza 30, Poland*

[3]AGH University of Science and Technology, Faculty of Physics and Applied Computer Science
*PL-30-059 Kraków, Av. A. Mickiewicza 30, Poland*

[4]Institute of Low Temperature and Structure Research, Polish Academy of Sciences
*PL-50-422 Wrocław, ul. Okólna 2, Poland*

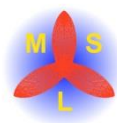

[*]Corresponding author: sfrueben@cyf-kr.edu.pl

## Abstract

$^{57}$Fe and $^{151}$Eu Mössbauer spectra were obtained versus temperature for Eu$_{0.57}$Ca$_{0.43}$Fe$_2$As$_2$ compound with 3d and 4f magnetic order and Eu$_{0.73}$Ca$_{0.27}$(Fe$_{0.87}$Co$_{0.13}$)$_2$As$_2$ re-entrant superconductor, where the finite resistivity reappears while approaching the ground state. They were compared with previously obtained spectra for parent compounds EuFe$_2$As$_2$ and CaFe$_2$As$_2$. It was found that substitution beyond the Fe-As layers does not lead to the rotation (canting) of the Eu$^{2+}$ magnetic moments and does not generate Eu$^{3+}$ states. On the other hand, re-entrant superconductor exhibits rotation (canting) of the Eu$^{2+}$ moments on the c-axis of the unit cell leading to the transferred hyperfine magnetic field on iron nuclei. Divalent europium orders magnetically within the bulk of the re-entrant superconducting phase. The re-entrant superconductor remains in the inhomogeneous state close to the ground state with about 27 % of the volume being free of 3d magnetism, while the remainder exhibits weak spin density wave. Those two regions slightly differ by the electric field gradient and electron density on iron nuclei.

## 1. Introduction

Iron-based superconductors of the '122' family are still very interesting compounds as they are a significant playground for competition between itinerant 3d magnetic order of the spin density wave (SDW) type (usually longitudinal and incommensurate) and superconductivity [1]. They could be prepared as relatively large single crystals of high purity and order free of other phases. Parent compounds of pnictides $AFe_2As_2$ (A=Ca, Sr, Ba, Eu) develop SDW accompanied by a slight lattice distortion from tetragonal to orthorhombic symmetry [2, 3]. For the A elements being rare earths with non-zero localized 4f magnetic moments one observes antiferromagnetic order of these moments at much lower temperatures [4]. Superconductivity could be achieved by suppression of the SDW order (and lattice distortion as well) by either external pressure [5, 6] or suitable doping on any site. One can apply electron doping, hole doping or iso-electronic substitution [7-9]. For the iso-electronic substitution to be successful one has to replace arsenic by e.g. phosphorus, i.e., one has to perturb significantly Fe-As layers with strong perturbation of the iron environment, as iron atoms are tetrahedrally coordinated by pnictogen [10]. It is interesting to note, that 4f antiferromagnetic order occurs within superconducting material as well, and doping leading to superconductivity usually causes some 4f magnetic moment reorientation (canting) with a generation of the 4f ferromagnetic component [11-14].

Mössbauer spectroscopy is very useful while looking at the peculiar magnetism of parent compounds [15-17] and iron-based superconductors [18-20]. The present contribution reports Mössbauer results obtained by means of $^{57}$Fe and $^{151}$Eu spectroscopy on the $Eu_{0.57}Ca_{0.43}Fe_2As_2$ compound being iso-electronic with compounds $EuFe_2As_2$ and $CaFe_2As_2$ [21]. Replacement of Eu by Ca does not perturb significantly Fe-As layer, and hence the $Eu_{0.57}Ca_{0.43}Fe_2As_2$ behaves like a parent compound with SDW magnetism and subsequent ordering of divalent Eu magnetic moments. On the other hand, the doubly substituted compound $Eu_{0.73}Ca_{0.27}(Fe_{0.87}Co_{0.13})_2As_2$ was found as re-entrant superconductor in similarity to the superconductor $Eu(Fe_{1-x}Co_x)_2As_2$, where superconductivity persists to the ground state [11]. Re-entrant behavior is observed here as a sharp transition to the superconducting state at about 12 K followed by the recovery to the normal state at about 2 K lower temperature. Doubly substituted compound is in the inhomogeneous state with a part of the volume developing weak SDW at low temperature. Recently a new '1144' family of iron-based stoichiometric superconductors has been discovered, e.g. $CaKFe_4As_4$ [22], based on the cell of the '122' family with layers separating Fe-As sheets being alternating and shifted Ca and K layers. These compounds do not exhibit 3d magnetism and crystallize in the orthorhombic symmetry [23].

## 2. Experimental

Single crystals of compounds were grown by Sn-flux method. The elements in respective molar ratios were loaded into alumina crucible and sealed in a silica tube under vacuum. The tube was heated slowly to 1050 °C and kept at this temperature for several hours, to dissolve all components. Then, it was cooled slowly at a rate of 2 °C/h to 650 °C. The liquid tin was decanted from the crucible at this stage. Residue of Sn was removed by etching in diluted hydrochloric acid.

The chemical composition of the grown single crystals was determined using EDS spectroscopy. The crystal structure and phase purity of the samples was characterized by powder X-ray diffraction using X'Pert Pro powder diffractometer equipped with a linear

PIXcel detector and CuKα radiation. Electrical resistivity was measured using standard four-probe technique in a Quantum Design PPMS platform. Magnetic susceptibility measurements show that magnetic ordering of divalent europium in $Eu_{0.57}Ca_{0.43}Fe_2As_2$ occurs at about 12 K in comparison with 19 K for $EuFe_2As_2$. Divalent europium orders at about 11 K for $Eu_{0.73}Ca_{0.27}(Fe_{0.87}Co_{0.13})_2As_2$ as shown by the magnetic susceptibility results. On the other hand, 3d itinerant magnetic order starts at somewhat higher temperature than for both parent compounds $EuFe_2As_2$ (192.1 K) and $CaFe_2As_2$ (175.3 K), respectively [21], as shown by resistivity measurements.

Mössbauer transmission measurements for 14.41-keV transition in $^{57}$Fe were performed using the RENON MsAa-3 spectrometer operated in the round-corner triangular mode and equipped with the LND Kr-filled proportional detector and He-Ne laser based interferometer used to calibrate a velocity scale. A single line commercial $^{57}$Co(Rh) source kept at room temperature was applied. The Mössbauer absorber for $Eu_{0.57}Ca_{0.43}Fe_2As_2$ was prepared using 40 mg of the material in the powder form and the absorber thickness amounted to 20 mg/cm$^2$ of investigated material. The absorber of the $Eu_{0.73}Ca_{0.27}(Fe_{0.87}Co_{0.13})_2As_2$ was prepared using 42 mg of the material in the powder form and the absorber thickness amounted to 21 mg/cm$^2$ of investigated material. The cryostat SVT-400 by Janis Research Inc. was used to maintain temperature of absorbers. $^{151}$Eu spectra for 21.6-keV resonant transition were collected applying $^{151}$SmF$_3$ source kept at room temperature and a scintillation detector. The WisseL spectrometer operated in the sinusoidal mode was used to collect spectra of the $Eu_{0.57}Ca_{0.43}Fe_2As_2$ and it was calibrated by means of the α-Fe absorber spectrum collected applying $^{57}$Co(Rh) source – both kept at room temperature. Temperature of the absorber was set and controlled by means of the closed cycle SHI-850-5 Janis Research Inc. He-refrigerator. Spectra of the $Eu_{0.73}Ca_{0.27}(Fe_{0.87}Co_{0.13})_2As_2$ compound were collected using He cryostat to cool the absorber and by means of the RENON MsAa-4 spectrometer calibrated with the He-Ne laser based interferometer. Absorbers were prepared as a mosaic of crystals to avoid oxidation. Data for $^{57}$Fe and $^{151}$Eu Mössbauer hyperfine parameters were processed by means of the Mosgraf-2009 software within the transmission integral approximation [24]. Spectral shifts are reported versus room temperature α-Fe or versus room temperature $^{151}$SmF$_3$ source, respectively.

3. **Results and discussion**

Figure 1 shows relative resistivity plotted versus temperature for three compounds of the composition $Eu_{1-x}Ca_xFe_2As_2$ and re-entrant superconductor $Eu_{0.73}Ca_{0.27}(Fe_{0.87}Co_{0.13})_2As_2$. Magnetic ordering of the 3d electrons is clearly seen for $Eu_{1-x}Ca_xFe_2As_2$ compounds with the highest ordering temperature being for the compound with intermediate composition. On the other hand, for the compound $Eu_{0.73}Ca_{0.27}(Fe_{0.87}Co_{0.13})_2As_2$ one can clearly see sharp transition from the metallic to the superconducting state at 12 K followed by re-entrant behavior below 10 K. A broad hump at lower temperatures seems typical for the gradual ordering of some magnetic moments.

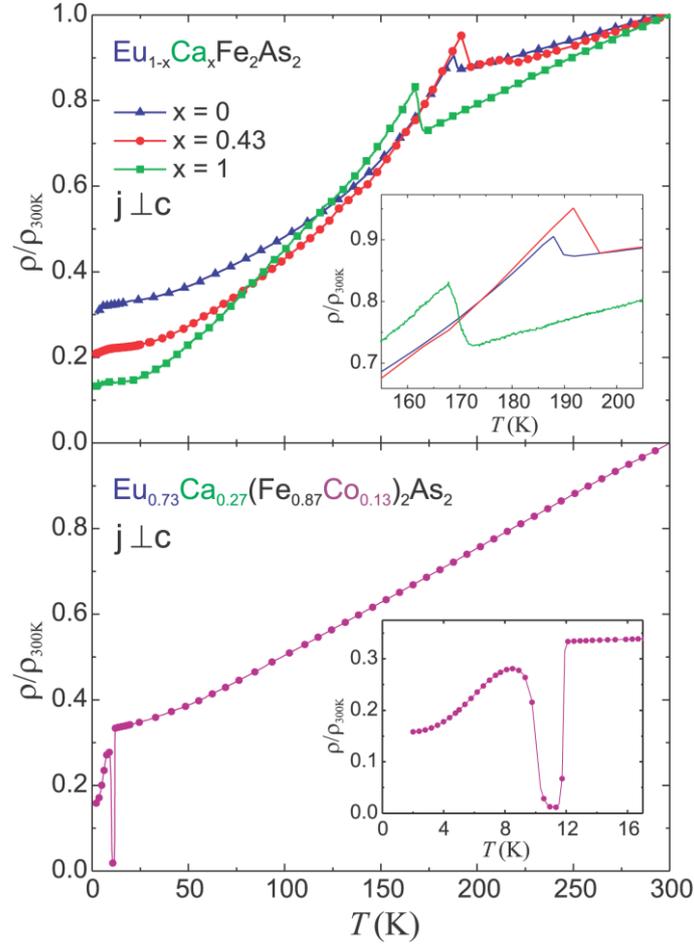

**Figure 1** Resistivity ρ normalized to resistivity at 300 K ($\rho_{300K}$) for compounds with well developed 3d magnetic order $Eu_{1-x}Ca_xFe_2As_2$ (x=0, 0.43, 1) and re-entrant superconductor $Eu_{0.73}Ca_{0.27}(Fe_{0.87}Co_{0.13})_2As_2$ plotted versus temperature $T$. The current **j** was applied perpendicular to the c-axis.

Figure 2 shows $^{57}$Fe Mössbauer transmission spectra for the compound $Eu_{0.57}Ca_{0.43}Fe_2As_2$ measured at various temperatures (central panel). Corresponding spectra for parent compounds $EuFe_2As_2$ (left panel) and $CaFe_2As_2$ (right panel) are shown as well. The symbol $S$ stands for the total central shift versus room temperature α-Fe, while the symbol $\Delta$ denotes electric quadrupole splitting (effective for the magnetically split spectra). The symbol $\sqrt{\langle B^2 \rangle}$ stands for the mean squared amplitude of the (incommensurate) spin density wave (SDW).

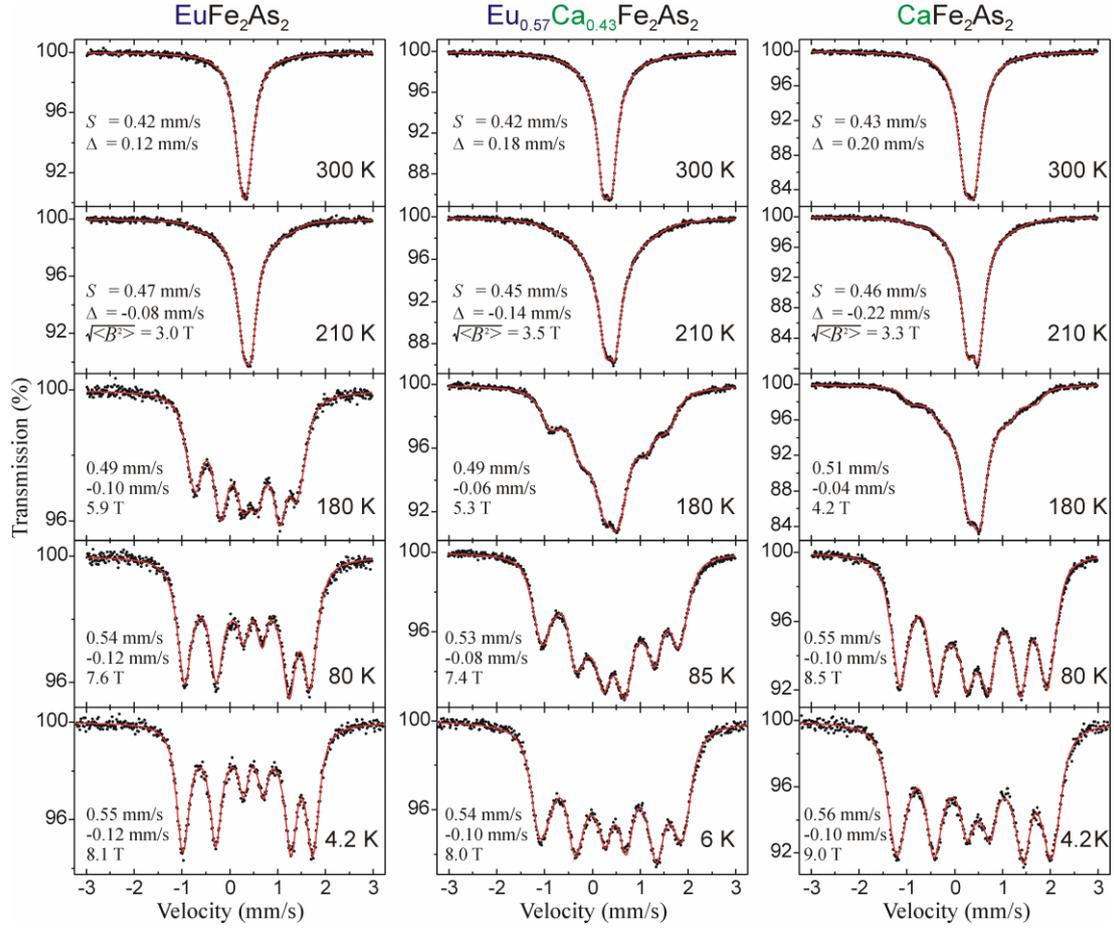

**Figure 2** $^{57}$Fe Mössbauer spectra of the Eu$_{1-x}$Ca$_x$Fe$_2$As$_2$ (x=0, 0.43, 1) compounds with well developed 3d magnetic order obtained versus temperature. Meaning of particular symbols is shown for the first two upper rows and remains the same for all rows.

Figure 3 shows shape of SDW denoted as $B(qx)$ versus phase shift $qx$ and corresponding normalized distribution of the hyperfine magnetic field (absolute value) $W(B)$ versus hyperfine magnetic field $B$ at 180 K and close to saturation for above compounds with well defined SDW. The symbol $B_{max}$ stands for the maximum amplitude of SDW, while symbols $h1$ and $h3$ denote amplitudes of the first and third harmonic of SDW. The symbol $\langle B \rangle$ stands for the average hyperfine field (absolute value).

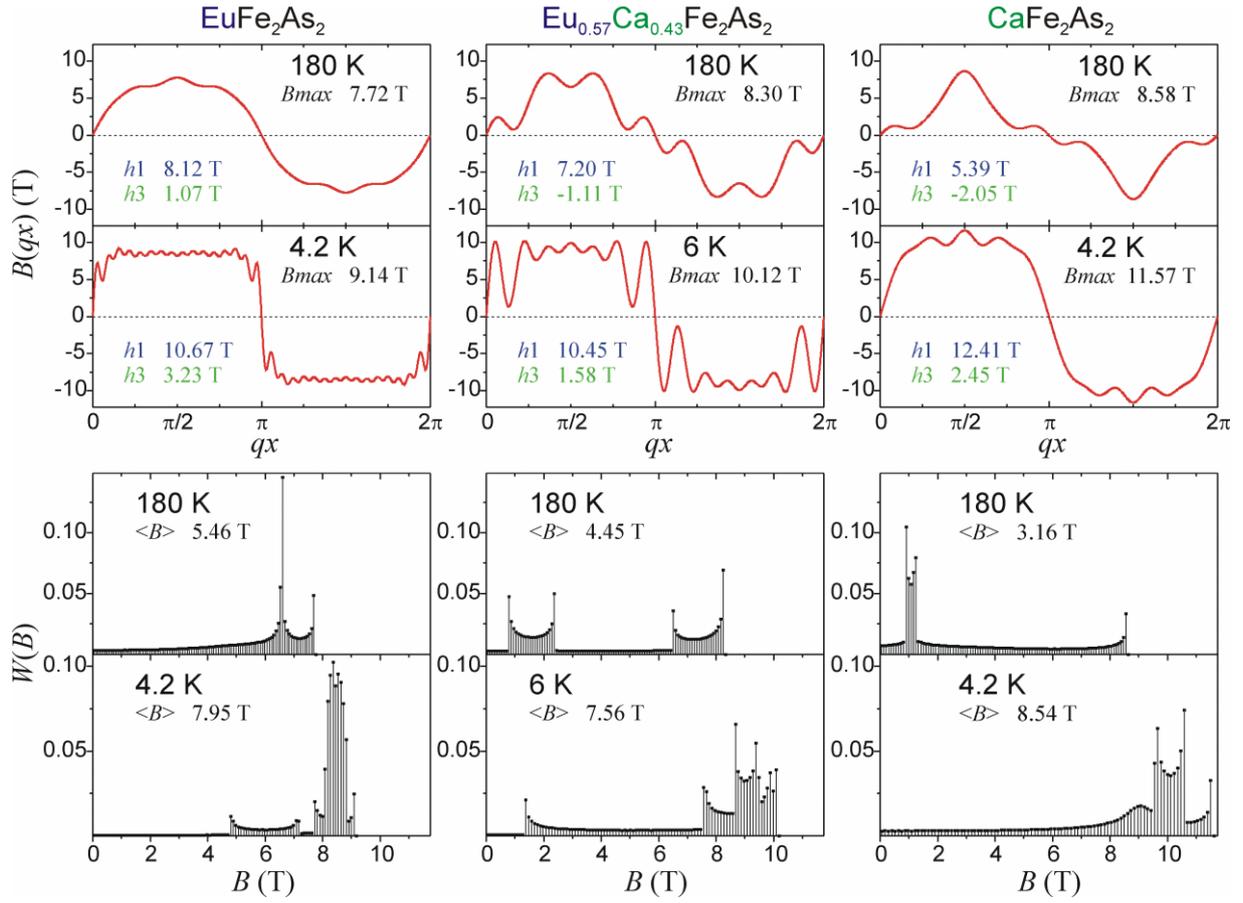

**Figure 3** Shapes of SDW and corresponding distributions of hyperfine magnetic fields at 180 K and close to magnetic saturation for $Eu_{1-x}Ca_xFe_2As_2$ (x=0, 0.43, 1) compounds.

More $^{57}$Fe Mössbauer spectra, corresponding SDW shapes and hyperfine magnetic field distributions for the compound $Eu_{0.57}Ca_{0.43}Fe_2As_2$ are shown versus temperature in Figure 4 below transition to the magnetic order of the SDW type.

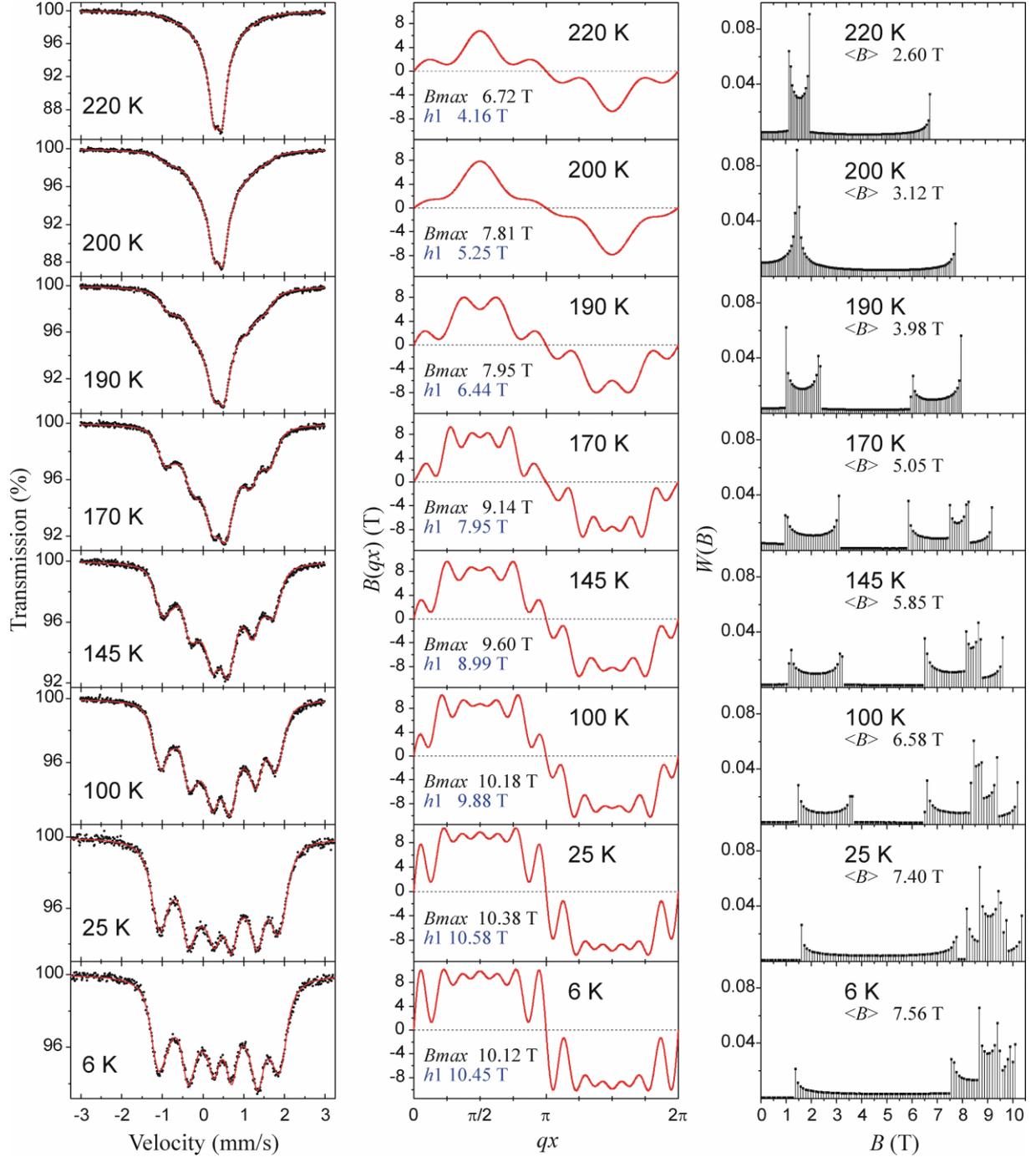

**Figure 4** $^{57}$Fe Mössbauer spectra, corresponding SDW shapes and hyperfine magnetic field distributions versus temperature for the Eu$_{0.57}$Ca$_{0.43}$Fe$_2$As$_2$ below transition to the SDW magnetic order.

Figure 5 summarizes essential results obtained for compounds with well developed SDW by means of $^{57}$Fe Mössbauer spectroscopy. Upper panel shows mean squared amplitude of SDW $\sqrt{\langle B^2 \rangle}$ versus temperature $T$. The symbol $T_{\text{SDW}}$ stands for the offset temperature of the SDW magnetic order (correlated with the orthorhombic distortion of the tetragonal unit cell upon lowering of the temperature). It is interesting to note that there is none transferred field on the iron nuclei due to the Eu$^{2+}$ magnetic order. Static critical exponents amount to 0.124 for EuFe$_2$As$_2$ and to 0.158 for compounds containing calcium. Lower panel shows total central shift $S$ versus temperature (relative to room temperature α-Fe). The variation seems entirely due to the second order Doppler shift (SOD) and yields Debye temperatures $\theta_D$ shown in Figure 5. Some small anomaly in the shift is observed at the temperature of 3d magnetic order (see, lower panel of Figure 5). It is likely to be correlated with the structural transition from the tetragonal to the orthorhombic phase accompanying magnetic transition.

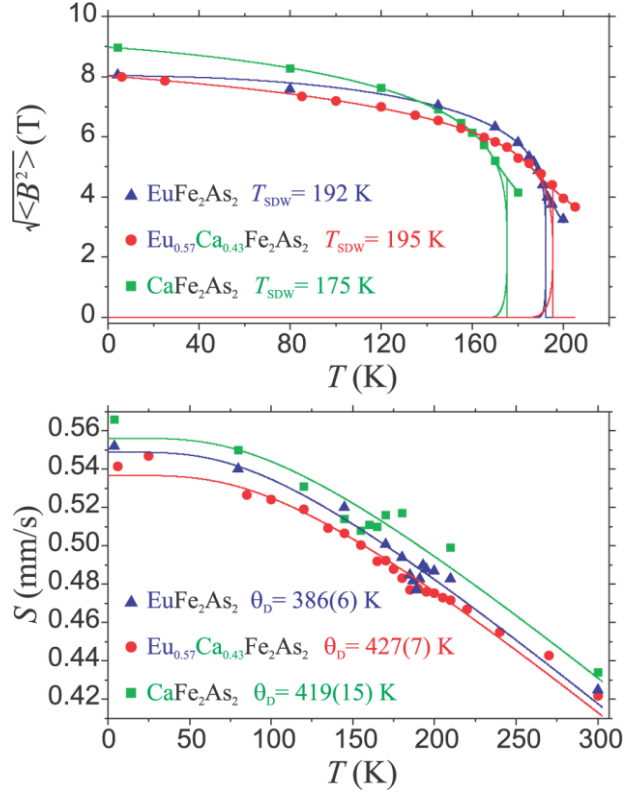

**Figure 5** Mean squared amplitudes of SDW and total central shift versus temperature for $^{57}$Fe Mössbauer spectra of the $Eu_{1-x}Ca_xFe_2As_2$ (x=0, 0.43, 1) compounds. Data of the upper panel are fitted according to the model described in Ref. [21] (solid lines), while data of the lower panel are fitted to the standard Debye model (solid lines). Typical errors for $\sqrt{\langle B^2 \rangle}$ and $S$ are 0.02 T and 0.002 mm/s respectively, so error bars are included in marked points.

$^{151}$Eu Mössbauer spectra of the compounds $EuFe_2As_2$, $Eu_{0.57}Ca_{0.43}Fe_2As_2$ and re-entrant superconductor $Eu_{0.73}Ca_{0.27}(Fe_{0.87}Co_{0.13})_2As_2$ are shown versus temperature in Figure 6. The symbol $S$ stands for the total central shift of the $Eu^{2+}$ component versus room temperature source made of $SmF_3$. The symbol $\varepsilon = \frac{1}{4}(c/E_0)eQ_gV_{zz}$ stands for the quadrupole coupling constant for $Eu^{2+}$ under assumption that the electric field gradient (EFG) is axially symmetric with the main axis aligned with the c-axis of the unit cell. The symbol $c$ stands for the speed of light in vacuum, the symbol $E_0$ denotes transition energy, while the symbol $e$ stands for the positive elementary charge. The symbol $Q_g$ denotes spectroscopic nuclear electric quadrupole moment in the ground state of $^{151}$Eu, while the symbol $V_{zz}$ stands for the principal component of EFG on $Eu^{2+}$ nucleus. The symbol $B$ denotes value of the hyperfine magnetic field on the $Eu^{2+}$ nucleus, while the symbol $\theta_c$ stands for the angle between principal component of EFG and hyperfine magnetic field on the $Eu^{2+}$ nucleus, i.e., for the angle between the c-axis and the hyperfine field. One can see that compounds with well developed

SDW do not contain non-magnetic (in the ground state) $Eu^{3+}$ ions. On the other hand, trivalent europium makes about 8 % contribution to the spectra of $Eu_{0.73}Ca_{0.27}(Fe_{0.87}Co_{0.13})_2As_2$ in similarity to the superconductor and/or overdoped $Eu(Fe_{1-x}Co_x)_2As_2$ (see, Ref. [11]). However, we were unable to fit the transferred field on $Eu^{3+}$ due to the magnetic order of $Eu^{2+}$ in contrast to the case of $Eu(Fe_{1-x}Co_x)_2As_2$ [11]. For trivalent europium one obtains $S = 0.7$ mm/s at all temperatures and $\varepsilon = -2$ mm/s above magnetic order of divalent europium. For 4.2 K one gets $\varepsilon = -6$ mm/s. Absorber linewidths for divalent europium amount to 1.5 mm/s above magnetic order of $Eu^{2+}$ and to 2.1 mm/s below. Corresponding linewidths for trivalent europium amount to 2 mm/s above mentioned order and to 3 mm/s below. Hence, one cannot easily proof that trivalent europium is in the same phase as remainder of the sample.

Divalent europium orders within the $Eu_{0.73}Ca_{0.27}(Fe_{0.87}Co_{0.13})_2As_2$ at sufficiently low temperature in similarity to the simpler case of $Eu(Fe_{1-x}Co_x)_2As_2$ superconductor or overdoped material [11]. The hyperfine magnetic field on the $Eu^{2+}$ ions in the spectra collected at lowest temperatures for calcium substituted compounds is slightly smaller than the field of 27.4(1) T in the parent $EuFe_2As_2$ [11]. Hence, a decrease is likely to be due to the magnetic dilution of the divalent Eu ions caused by Ca-substitution. The angle $\theta_c$ proofs that antiferromagnetic order of divalent europium between subsequent a-b planes is preserved as long as Fe-As layers remain unperturbed. On the other hand, for $Eu_{0.73}Ca_{0.27}(Fe_{0.87}Co_{0.13})_2As_2$ the angle between hyperfine field on $Eu^{2+}$ and the main component of EFG on the same ion amounts to 30(4)° at 4.2 K. The main component of EFG is oriented along the c-axis, while the hyperfine field is aligned with the $Eu^{2+}$ magnetic moment. Hence, the magnetic moment of divalent europium is tilted by 30(4)° from the c-axis, while in the parent compounds is perpendicular to the c-axis [11]. In $Eu_{0.73}Ca_{0.27}(Fe_{0.87}Co_{0.13})_2As_2$ magnetic moments of $Eu^{2+}$ tend to rotate on the c-axis in similarity to the superconducting or overdoped compound $Eu(Fe_{1-x}Co_x)_2As_2$ with some canting leading to the ferromagnetic component of the 4f origin [11]. For comparison in superconducting $Eu(Fe_{0.82}Co_{0.18})_2As_2$ with similar Co concentration one has $\theta_c = 44(1)°$ at 4.2 K, while for the overdoped $Eu(Fe_{0.71}Co_{0.29})_2As_2$ one gets $\theta_c = 29(3)°$ at 4.2 K [11].

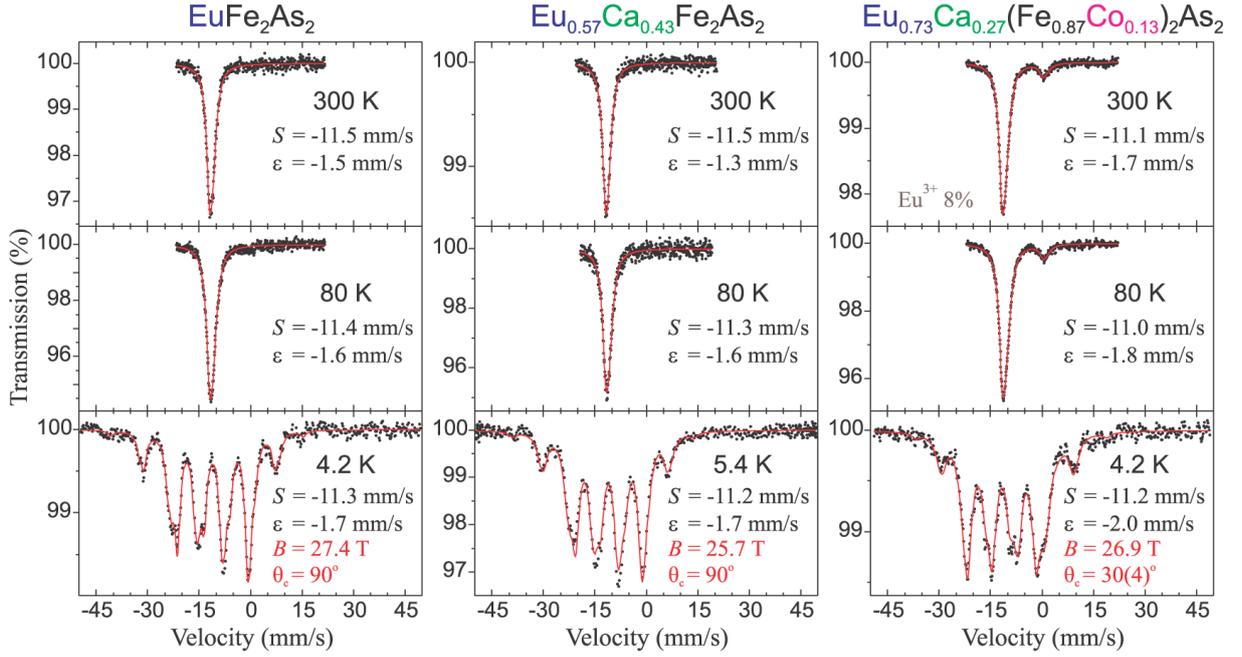

**Figure 6** $^{151}$Eu Mössbauer spectra of europium-based compounds having well developed SDW and re-entrant superconductor $Eu_{0.73}Ca_{0.27}(Fe_{0.87}Co_{0.13})_2As_2$ versus temperature. The symbol *S* stands for the spectral shift versus room temperature $SmF_3$ source. The symbol $\varepsilon$ denotes quadrupole coupling constant. The hyperfine field on $Eu^{2+}$ is denoted by *B*, while the symbol $\theta_c$ stands for the angle between c-axis of the unit cell and magnetic moment of the $Eu^{2+}$.

$^{57}$Fe Mössbauer spectra of the re-entrant superconductor $Eu_{0.73}Ca_{0.27}(Fe_{0.87}Co_{0.13})_2As_2$ are shown in Figure 7 versus temperature. A transition to the superconducting state occurs at $T_{sc} = 12$ K. Spectra consist of two electric quadrupole split components at high temperature. The minor component is characterized by very broad lines. The major component has slightly larger spectral shift (isomer shift) and much larger quadrupole splitting at room temperature than the parent compound $EuFe_2As_2$ [21]. Hence, the electron density on iron nuclei is lowered due to doping, while the increase in the quadrupole splitting is likely to be caused by the chemical disorder induced by dopants. Traces of SDW order appear in the spectrum collected at 80 K. About 44 % of the sample volume contains SDW with the mean squared amplitude of about 2.7 T, while remainder is free of the iron induced 3d magnetism at 80 K. Spectrum obtained close to the ground state, i.e., at 4.2 K consists of two components as well. About 73 % of the sample volume exhibits SDW and some hyperfine field transferred from the ordered $Eu^{2+}$ magnetic moments. Remainder of the iron nuclei experiences a transferred hyperfine field of about 1.1 T due to the divalent europium magnetic order. The latter minor component does not exhibit 3d magnetic order and it is responsible for filamentary superconductivity observed for 'Eu-122' iron-based superconductors [11].

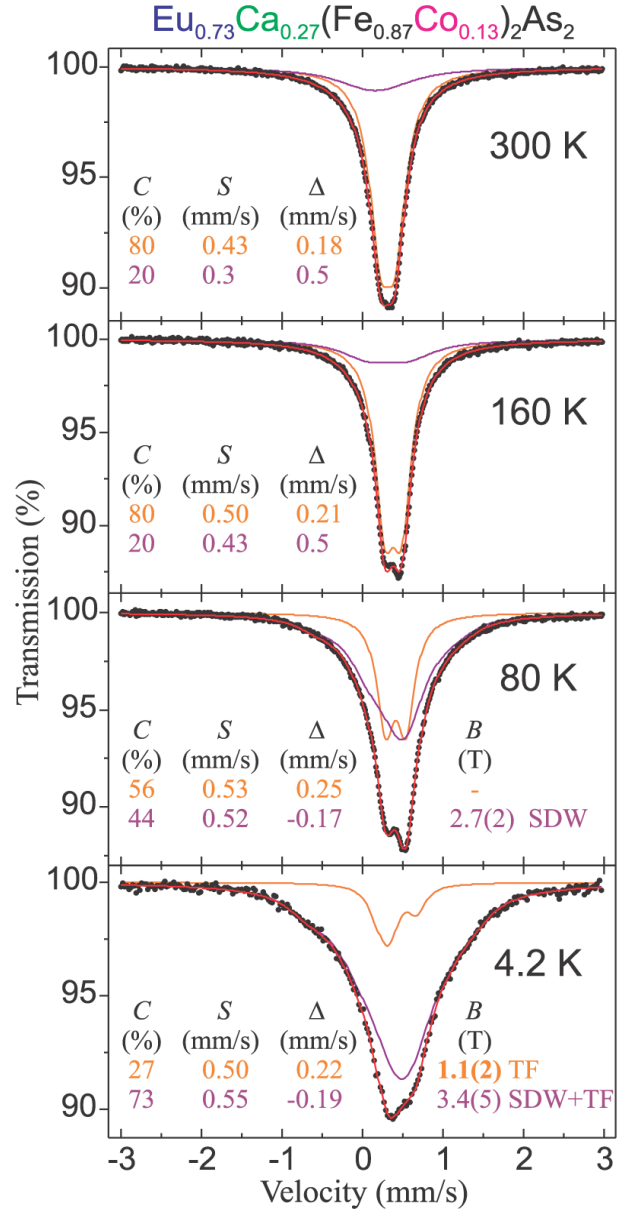

**Figure 7** $^{57}$Fe Mössbauer spectra of re-entrant superconductor $Eu_{0.73}Ca_{0.27}(Fe_{0.87}Co_{0.13})_2As_2$ with $T_{sc} = 12$ K and essential Mössbauer spectroscopy parameters. The symbols denote: $C$ - relative contribution to the absorption cross-section, $S$ - total spectral shift versus room temperature α-Fe, $\Delta$ - electric quadrupole splitting (effective for the magnetically split spectra), $B$ - magnetic hyperfine field. Errors are of the order of unity for the last digit shown or as stated. TF stands for the transferred field from magnetically ordered $Eu^{2+}$ on the iron nucleus. SDW denotes spin density wave on Fe. Respective sub-components are shown.

## 4. Conclusions

The SDW develops in a similar fashion for the compound $Eu_{0.57}Ca_{0.43}Fe_2As_2$ as for parents $EuFe_2As_2$ and $CaFe_2As_2$ with transition temperature being somewhat higher than for both parent compounds [21]. However, the shape of SDW is less regular and one can observe regions of high and low hyperfine magnetic field (itinerant 3d magnetic moment) within approximate temperature range 100 – 190 K. However, when comparing SDW shapes derived from the spectra measured on polycrystalline samples one has to remember that they sensitively depend on the size of grains [25-27].

The static critical exponent for $Eu_{0.57}Ca_{0.43}Fe_2As_2$ compound is the same as for Ca-based parent compound being slightly higher than for the Eu-based parent compound. All those exponents indicate itinerant almost two-dimensional magnetic order of the 3d type. Addition of calcium does not generate trivalent europium. On the other hand, divalent europium orders antiferromagnetically between subsequent a-b planes at somewhat lower temperature than for the parent compound. Hence, no transferred magnetic field due to divalent europium magnetic order appears on iron nuclei in similarity to the Eu-based parent compound.

The re-entrant superconductor $Eu_{0.73}Ca_{0.27}(Fe_{0.87}Co_{0.13})_2As_2$ exhibits suppression of SDW with about 27 % of the volume being free of 3d magnetic order close to the ground state. These regions seem to be responsible for superconductivity. Some trivalent europium appears, but it is impossible to proof that it belongs to the same phase in contrast to the superconductor and/or overdoped $Eu(Fe_{1-x}Co_x)_2As_2$ compound [11]. It is known that trivalent europium is generated by the local pressure due to the chemical substitution [28] or being external [29]. Divalent europium orders antiferromagnetically at low temperature with a significant magnetic moment component parallel to the c-axis and having some ferromagnetic character. This situation is very similar to the situation encountered in the compound $Eu(Fe_{1-x}Co_x)_2As_2$ and leads to the transferred magnetic field on iron nuclei [11]. Magnetic order of divalent europium occurs within re-entrant phase as well as for superconducting phase $Eu(Fe_{1-x}Co_x)_2As_2$ and it has very similar character in both cases [11]. It is interesting to note, that for doubly substituted compound electron density and EFG on iron nuclei are somewhat different in the re-entrant superconducting regions in comparison with SDW regions.

It is likely that somewhat strange re-entrant behavior for $Eu_{0.73}Ca_{0.27}(Fe_{0.87}Co_{0.13})_2As_2$ compound is due to the divalent europium magnetic ordering. Random substitution of europium by calcium (without cobalt substitution) does not generate 4f component of the magnetic moment along the c-axis. On the other hand, it generates randomly fluctuating ferromagnetic component along the a-axis due to lack of 4f magnetic partners for some $Eu^{2+}$ ions. Subsequent substitution of cobalt leads to the 4f magnetic moment rotation on the c-axis with subsequent ferromagnetic component along this axis. This component is randomly enhanced by the presence of calcium due to the same mechanism as described above. The 4f magnetic order of any kind is unlikely to break Cooper pairs within Fe-As layers. Hence, one can expect for doubly substituted compounds a development of some coupling between 4f moments and 3d itinerant spins leading to the high magnetic polarizability of the otherwise "very hard" 3d spins. Such polarized states despite being disordered could break Cooper pairs, while driven indirectly by the order of 4f moments and contribute to the resistivity in the manner observed. Another possibility is a development of the pair-breaking Andreev scattering on the borders between SDW regions and regions free of 3d magnetic order. Necessary ferromagnetic component could be again provided by the divalent europium magnetic ordering with the symmetry broken by a more or less random substitution of europium by calcium.